\documentclass[11pt]{article}
 \usepackage{graphicx}

\usepackage{amsmath}
\usepackage{amssymb}
\usepackage{amsmath}
\usepackage{amssymb}
\usepackage{amscd}
\newtheorem{Definition}{Definition}[section]

\newtheorem{Theorem}[Definition]{Theorem}

\topmargin = - 1 cm \textheight = 23 cm \textwidth = 15 cm
\newcommand{\epr}{{\sc epr}}
\renewcommand{\L}{\label} 

\newcommand{\half}{\mbox{\footnotesize $\frac{1}{2}$}}
\newcommand{\tthird}{\mbox{\footnotesize $\frac{2}{3}$}}
\newcommand{\third}{\mbox{\footnotesize $\frac{1}{3}$}}
\newcommand{\beq}{\begin{equation}}
\newcommand{\eeq}{\end{equation}} 
\newcommand{\bea}{\begin{eqnarray}}
\newcommand{\eea}{\end{eqnarray}}

\newcommand{\qm}{quantum mechanics}

 \newcommand{\til}{\tilde}
\newcommand{\raw}{\rightarrow}

 \newcommand{\Raw}{\Rightarrow}

\newcommand{\la}{\langle} \newcommand{\ra}{\rangle}

\newcommand{\x}{\times} 
 
\newcommand{\er}{\eqref}
\newcommand{\al}{\alpha}

\newcommand{\lm}{\lambda} 
 
\newcommand{\Sg}{\Sigma}  
 
\newcommand{\ch}{\ch}  
 
\renewcommand{\L}{\label}

\newcommand{\C}{{\mathbb C}} 
 
 \newcommand{\R}{{\mathbb R}}
 
\newcommand{\ba}{\mathbf{a}}\newcommand{\bb}{\mathbf{b}}

  \makeatletter
\newskip\tempskip \def\endproof{{\parfillskip24\p@ plus\@ne
fil\@@par}\tempskip\prevdepth
\ifdim\lastskip=\z@\tempskip\z@\else\vskip-\lastskip
\ifdim\tempskip>4\p@ \tempskip.5\tempskip \else \tempskip\z@\fi\fi
\nobreak\vskip-\baselineskip\vskip-\tempskip\noindent\hbox
to\hsize{\hfill
$\blacksquare$}\par\vskip\tempskip\vskip\abovedisplayskip\@doendpe}
\makeatother \makeatletter
\newskip\tempskip \def\endiproof{{\parfillskip24\p@ plus\@ne
fil\@@par}\tempskip\prevdepth
\ifdim\lastskip=\z@\tempskip\z@\else\vskip-\lastskip
\ifdim\tempskip>4\p@ \tempskip.5\tempskip \else \tempskip\z@\fi\fi
\nobreak\vskip-\baselineskip\vskip-\tempskip\noindent\hbox
to\hsize{\hfill
$\Box$}\par\vskip\tempskip\vskip\abovedisplayskip\@doendpe}
\makeatother \newcommand{\enp}{\endproof}

\begin{document} 
\pagenumbering{arabic} \setlength{\unitlength}{1cm}\cleardoublepage
\thispagestyle{empty}
\title{Constraints on determinism: Bell versus Conway--Kochen\footnote{Dedicated to Professor Hans Maassen, on the occasion of his inaugural lecture (15-01-2014).}}
\author{Eric Cator and Klaas Landsman\\ \mbox{} \hfill \\
Institute for Mathematics, Astrophysics, and Particle Physics\\ Faculty of Science, Radboud University Nijmegen\\
\texttt{e.cator@science.ru.nl, landsman@math.ru.nl}}
\date{\today}
\maketitle\vspace{-1cm}
 \begin{abstract} 
\noindent 
Bell's Theorem from 1964 and the (Strong) Free Will Theorem of Conway and Kochen from 2009 both exclude deterministic hidden variable theories (or, in modern parlance, `ontological models') 
that are compatible with some small fragment of \qm, admit `free' settings of the archetypal Alice\&Bob experiment, and satisfy a locality condition akin to Parameter Independence. We clarify the relationship between these theorems by giving reformulations 
of both that exactly pinpoint their resemblance and their differences.  Our reformulation imposes  determinism in what we see as the only  consistent way, in which the `ontological state'  initially determines both the settings and the outcome of the experiment. The usual status of the settings as `free' parameters is subsequently  recovered from  independence assumptions on the pertinent (random) variables. Our reformulation also clarifies the role of the settings in 
Bell's later generalization of his theorem to stochastic hidden variable theories.

\end{abstract}
\maketitle
\section{Introduction}

Though not really new,\footnote{Analogous earlier results were obtained, in chronological order, by Heywood \& Redhead \cite{HR}, Stairs \cite{Stairs}, Brown \& Svetlichny \cite{BS}, and Clifton \cite{Clifton} (of which only \cite{HR} was cited by Conway and Kochen).}  the (Strong) Free Will Theorem of Conway and Kochen  \cite{CK1,CK2} is one of the sharpest and most interesting results that give constraints on determinism. It does so by  proving that determinism is incompatible with a number of \emph{a priori} desirable assumptions, including a small fragment of \qm\ (viz.\ the theory of two \epr-correlated spin-one particles), the free choice of settings of an \epr-style bipartite experiment involving such particles, and a  locality 
condition called {\sc min}. The latter has a long pedigree, arguably going back to {\sc epr}, but it was first stated quite clearly by Bell:\footnote{Bell \cite{Bell1} even attributes it to Einstein.
See \cite{Wiseman} for a detailed  analysis of the way this condition is actually used by Bell in \cite{Bell1,Bell3},
and of the way it has been (mis)perceived by others. In particular, one should distinguish it
from the locality condition usually named after Bell \cite{Bell4}. The latter, also called \emph{local causality},  is a conjunction of  two (probabilistic) notions  that are now generally called \emph{Parameter Independence}  ({\sc pi})
and \emph{Outcome Independence} ({\sc oi}); see  \cite{Bub,Jaeger,Jarrett,Maudlin,Seevinck,Shimony}.
The latter is automatically satisfied in the type of  deterministic  theories studied in \cite{Bell1,CK1,CK2}, upon which the former reduces to the condition  stated in the main text above, but now \emph{conditioned on certain values of the hidden variables}. Note that our  definition of the term {\sc pi} will  be different from the literature so far, though in the same spirit.} 
\begin{quotation}
`The vital assumption is that the result $B$ for particle 2 does not depend on the setting $\vec{a}$ of the magnet for particle 1, nor $A$ on $\vec{b}$.' \cite[p.\ 196]{Bell1}.
\end{quotation} 
In any case, also a closer study shows that Bell's (1964) Theorem on deterministic hidden variable theories  and the (Strong) Free Will Theorem appear to achieve a very similar (if not identical)  goal under strikingly similar assumptions, which prompts the question what exactly their mutual relationship is. 
Curiously, despite the stellar fame of Bell's 1964 paper (which according to Google Scholar had about 8500 citations as of May 2014) and the considerable attention that also the Free Will Theorem has received (e.g., \cite{GT,GtH}), as far as we are aware, there has been little research in this precise direction.\footnote{The only significant exception we could find is the small and otherwise interesting book by Hemmick and Shakur \cite{HS}, whose scathing  treatment of the  Free Will Theorem  is somewhat undermined by their claim (p.\ 90) that the assumption of determinism follows from the other assumptions in the Strong Free Will Theorem (notably {\sc pi} and perfect correlation). This seems questionable \cite{Wiseman}: 
either Bell's (later) locality condition  (i.e., {\sc pi} \emph{plus} {\sc oi})  in conjunction with perfect correlation  implies determinism,
or   {\sc pi} plus  determinism  implies {\sc oi} (and hence Bell Locality). Perhaps our view (which is certainly shared by Conway and Kochen!) that the assumptions of the Strong Free Will Theorem have been chosen quite carefully is clearer from our reformulation below than from even their second paper \cite{CK2} (not to speak of  their first \cite{CK1}). Indeed, if valid, the objection of  Hemmick and Shakur could just as well be raised against Bell's 1964 Theorem, where it would be equally misguided if both results are construed as attempts to put constraints on determinism in the first place. Our treatment of parameter settings will also be different from \cite{HS}.} 

Hence the main aim of this paper is to clarify the relationship between Bell's 1964 Theorem and the Free Will Theorem. But in doing so, we will \emph{en passant} attempt to resolve an issue that has troubled Bell as well as Conway and Kochen, namely the theoretical status of parameter settings.
As pointed out by Conway and Kochen themselves \cite{CK2}, it is odd to assume determinism for the physical system under consideration but not for the experimenters, so that the contradiction that proves their theorem seems almost circular.\footnote{This even led them to their curious way of paraphrasing their theorem as showing that 
`If we humans have free will, then elementary particles already
have their own small share of this valuable commodity'.}

 In  Bell's later work, there has  been a similar tension between the idea that the hidden variables (in the pertinent causal past) should on the one hand include all ontological information relevant to the experiment, but on the other hand should leave Alice and Bob free to choose any settings they like; see especially \cite{Norsen1,SU} for a fine analysis of Bell's  dilemma (by some of his greatest supporters).\footnote{See also \cite{Bell4,JB} and most recently \cite{Leifer} for the interpretation of hidden variables as ontological states.}
 We will show that in both contexts of Bells' Theorem (i.e.\ either deterministic or stochastic) this issue can be resolved in a straightforward way  by initially including the settings among the random variables describing the experiment, after which they are `liberated'  by suitable independence assumptions.\footnote{See also Colbeck and Renner \cite{CR1} for at least the first step of this strategy in the context of stochastic hidden variable theories. Using settings as labels, on the other hand, is defended in e.g.\ \cite{Bub,JB,SU}.}
  
The plan of our paper is as follows. In  Section \ref{BTsec} we present a version of Bell's original (1964) theorem \cite{Bell1} that addresses the above issues. As a warm-up for what is to come, in Section  \ref{BT1sec} we extend this version to the spin-one case, followed  in Section \ref{FWTsec} by a reformulation of the Strong Free Will Theorem \cite{CK2} in the same spirit. Our final Section \ref{BLTsec} goes beyond our primary goal of finding constraints on determinism, but has been included in order to show that our treatment of parameter settings through random variables also applies to Bell's later results on stochastic hidden variable theories \cite{Bell4,Bub,JB,Gill,Jaeger,Maudlin,Seevinck,Wiseman}.

Our conclusion is  that the Strong Free Will Theorem uses fewer assumptions than Bell's 1964 Theorem, as no appeal to probability theory is made.  This comes at a price, though. First, in the absence of an Aspect-type experiment using spin-one particles, the former so far lacks the experimental backing of the latter. Second, because of its dependence  on the Kochen--Specker Theorem, the Strong Free Will Theorem might lack finite precision robustness, cf.\  \cite{Appleby,BK,Hermens}, though this threat recently seems to have been obviated \cite{Hermens2}.
\section{Bell's (1964) Theorem revisited}\label{BTsec}
The setting of Bell's Theorem in its simplest form is given by the usual \epr-Bohm experiment (with photons) \cite{Bub}, in which Alice and Bob each choose a setting $A=\al\in X_A$ and $B=\beta\in X_B$, respectively, where $X_A$ and $X_B$ are finite sets whose elements are angles  in $[0,\pi)$. For the theorem, it is even enough to assume $X_A=\{\al_1,\al_2\}$ and $X_B=\{\beta_1,\beta_2\}$, for suitable $\al_i$ and $\beta_j$ (see below). Alice and Bob each receive one photon from an \epr-correlated pair, and determine whether or not it passes through a polarizer whose principal axis is set at an angle $\al$ or $\beta$ relative to some reference axis in the plane orthogonal to the direction of motion of the photon pair.\footnote{Equivalently, $\al$ and $\beta$ could stand for the corresponding unit vectors $\vec{a}$ and $\vec{b}$, defined up to a  sign.} If Alice's photon passes through she writes down $F=1|A=\al$, and if not she writes $F=0|A=\al$. Likewise, Bob records his result as $G=1|B=\beta$ or $G=0|B=\beta$. Repeating this experiment, they determine empirical probabilities $P_E$ for all possible outcomes through the frequency interpretation of probabilities, which they denote by $P_E(F=\lm|A=\al)$ and $P_E(F=\mu|B=\beta)$,
or, having got together and compared their results, by $P_E(F=\lm,G=\mu|A=\al,B=\beta)$, where $\lm,\mu\in\{0,1\}$. 
If the photon pair is prepared in the \epr-correlated state $|\psi_{\mbox{\sc epr}}\ra=(|0\ra|0\ra+|1\ra|1\ra)/\sqrt{2}$ (taking into account  helicity only), they find (as confirmed by \qm):\footnote{\label{fn7} Here $P_E(F\neq G|A=\al,B=\beta)\equiv P_E(F=0,G=1|A=\al,B=\beta)+P_E(F=1,G=0|A=\al,B=\beta)$. The complete statistics are: $P_E(F=1,G=1|A=\al,B=\beta)=P_E(F=0,G=0|A=\al,B=\beta)=\half \cos^2(\al-\beta)$ and $P_E(F=0,G=1|A=\al,B=\beta)=P_E(F=1,G=0|A=\al,B=\beta)=\half \sin^2(\al-\beta)$.} 
\begin{equation}
P_E(F\neq G|A=\al,B=\beta)=\sin^2(\al-\beta).\label{uitkomstEPR}
\end{equation}

The question, then, is whether these probabilities are `intrinsic' or `irreducible', as claimed by mainstream \qm, or instead are just a consequence of our ignorance. To make this precise, we  define the latter case, i.e.\ determinism, at least in our present context, adding the other assumptions of Bell's (1964) Theorem along the way.
\begin{Definition}\label{Belldefs}
In the context of the  \epr-Bohm experiment (with photons):
\begin{itemize}
\item 
\textbf{Determinism} means that there is a state space $X$  with associated functions 
\begin{equation}
A: X\raw X_A, B:X\raw X_B, F:X\raw\{0,1\}, G:X\raw\{0,1\},\label{ABFG}
\end{equation}
which completely describe the experiment in the sense that some state $x\in X$ determines \emph{both} its settings $\al=A(x),\beta=B(x)$ \emph{and} its outcome $\lm=F(x),\mu=G(x)$. 
\item
\textbf{Probability Theory} means that the above set $X$ can be upgraded to a probability space $(X,\Sg,P)$, carrying the above functions $A,B,F,G$ as random variables,\footnote{This formulation incorporates the assumption that $P$ is independent of $A,B,F,G$, and \emph{vice versa}. }
 so that the empirical  probabilities are reproduced as conditional joint probabilities through\footnote{Here 
$P(F=\lm,G=\mu|A=\al,B=\beta)\equiv P(F=\lm,G=\mu,A=\al,B=\beta)/P(A=\al,B=\beta)$ and 
$P(F=\lm,G=\mu,A=\al,B=\beta)\equiv P(\{x\in X\mid F(x)=\lm,G(x)=\mu,A(x)=\al,B(x)=\beta\})$, etc.}
\begin{equation}
P_E(F=\lm,G=\mu|A=\al,B=\beta)= P(F=\lm,G=\mu|A=\al,B=\beta). \label{PEP}
\end{equation}
Furthermore, in terms of a postulated additional random variable $Z:X\raw X_Z$:
\item \textbf{Parameter Independence}  means   that
$F=F(A,Z)$ and $G=G(B,Z)$,
 in  that there are measurable functions $\hat{F}:X_A \x X_Z\raw \{0,1\}$ and $\hat{G}:X_B\x X_Z\raw \{0,1\}$
  for which  $F(x)=\hat{F}(A(x),Z(x))$ and $G(x)=\hat{G}(B(x),Z(x))$ ($P$-almost everywhere).
\item \textbf{Freedom} means that
 $(A,B,Z)$ are probabilistically independent relative to  $P$.\footnote{On the usual definition, this also implies that the pairs $(A,B)$, $(A,Z)$, and $(B,Z)$ are independent.}
\end{itemize}
\end{Definition}
\newpage
\noindent Here $Z$ is the traditional `hidden variable' space that, in the spirit of Bell \cite{Bell4,Norsen1,SU},
carries exactly the  `ontological' information (including e.g.\ the photon variables) that is:
\begin{description}
\item i)  sufficiently complete for the outcome of the experiment to depend on $(A,B,Z)$ alone; 
\item  ii) independent of the settings $(A,B)$, in the pertinent probabilistic sense.
\end{description}
These conditions stand (or fall) together: without ii), i.e., \emph{Freedom}, one could  take $X_Z=X$ and $Z=\mathrm{id}$, whereas without i), $X_Z$ could be a singleton. 
\emph{Parameter Independence} in fact  sharpens i), which \emph{a priori} might have been $F=F(A,B,Z)$ and $G=G(A,B,Z)$,  to the effect that Alice's outcome is independent of Bob's, given $A$ and $Z$ (and \emph{vice versa}) \cite{Bell1}.
\smallskip

Our reformulation of Bell's (1964) Theorem, then, is as follows.
\begin{Theorem}\label{Bell1964}
Determinism, Probability Theory,  Parameter Independence, Freedom, and Nature (i.e.\ the outcome \er{uitkomstEPR} of the \epr-Bohm experiment) are contradictory.
\end{Theorem}
\emph{Proof.} Determinism, Probability Theory, and Parameter Independence imply\footnote{This is true even if $F=F(A,B,Z)$ and $G=G(A,B,Z)$ rather than $F=F(A,Z)$ and $G=G(B,Z)$.}
\begin{equation}
P(F=\lm,G=\mu|A=\al,B=\beta)=P_{ABZ}(\hat{F}=\lm,\hat{G}=\mu|\hat{A}=\al,\hat{B}=\beta),\label{norhat}
\end{equation}
where the function $\hat{A}:X_A \x X_B\x X_Z\raw X_A$ is just projection on the first coordinate, likewise the function $\hat{B}: X_A \x X_B\x X_Z\raw X_B$ is  projection on the second, and 
 $P_{ABZ}$ is the joint probability on $X_A \x X_B\x X_Z$ induced by the triple $(A,B,Z)$ and the probability measure $P$.
 Similarly, let $P_Z$ be the probability on $X_Z$ defined by $Z$ and $P$, and 
  define the following random variables on the probability space $(X_Z,\Sg_Z,P_Z)$:
\begin{eqnarray}
\hat{F}_{\al}(z)&:=& \hat{F}(\al,z);\label{hatFalbeta}\\
\hat{G}_{\beta}(z)&:=& \hat{G}(\beta,z).\label{hatGalbeta}
\end{eqnarray}
 Freedom then implies (indeed, is equivalent to the fact) that  $P_{ABZ}$ is given by a product measure on  $X_A \x X_B\x X_Z$ (cf.\ \cite[Lemma 3.10]{Kallenberg}). A brief computation then yields
  \begin{equation}
P_{ABZ}(\hat{F}=\lm,\hat{G}=\mu|\hat{A}=\al,\hat{B}=\beta)=P_Z(\hat{F}_{\al}=\lm,\hat{G}_{\beta}=\mu),
\end{equation}
and hence, from \er{norhat}, 
 \begin{equation}
 P(F=\lm,G=\mu|A=\al,B=\beta)=P_Z(\hat{F}_{\al}=\lm,\hat{G}_{\beta}=\mu).\label{crux2bis}
\end{equation}
Adding the \emph{Nature} assumption, i.e.\ \er{uitkomstEPR},  then gives the crucial result
\begin{equation}
P_Z(\hat{F}_{\al}\neq \hat{G}_{\beta})=\sin^2(\al-\beta).\label{uitkomstEPRhat}
\end{equation}
However, any four $\{0,1\}$-valued random variables must satisfy the (`Boole') inequality \cite{Pitowsky}
\beq P_Z(\hat{F}_{\al_1}\neq \hat{G}_{\beta_1})\leq P_Z(\hat{F}_{\al_1}\neq \hat{G}_{\beta_2})+P_Z(\hat{F}_{\al_2}\neq \hat{G}_{\beta_1})+P_Z(\hat{F}_{\al_2}\neq \hat{G}_{\beta_2}), \L{bell4tris} \eeq
which can be  proved directly from the axioms of (classical) probability theory. But for  suitable values of $(\al_1,\al_2,\beta_1,\beta_2)$ this inequality is violated by \er{uitkomstEPRhat}. Take, for example,  $\al_2=\beta_2=3\theta$, $\al_1=0$, and $\beta_1=\theta$. The inequality \er{bell4tris} then assumes the form $f(\theta)\geq 0$ for $f(\theta)=\sin^2(3\theta)+\sin^2(2\theta)-\sin^2(\theta)$. But this is false for many values of $\theta\in[0,2\pi]$.
 \enp
 
As already mentioned, in the usual treatment of Bell's Theorem (either his deterministic version \cite{Bell1,Bub} or his  stochastic version \cite{Bell4,Bub,JB,Gill,Jaeger,Maudlin,Seevinck,WW}), the hidden variable $\lm$ corresponds to  our
 $z\in X_Z$ rather than $x\in X$. It is the distinction  between $X_Z$ and the `super-deterministic' state space $X$ that allowed us to give a consistent formulation of \emph{Determinism} without jeopardizing  \emph{Freedom}. As  shown above, this eventually enables one to treat the apparatus settings as parameters rather than as random variables.

\section{Bell's (1964) Theorem for spin-one}\label{BT1sec}
The Free Will Theorem relies on a variation of the \epr-Bohm experiment in which $\C^2$ is replaced by $\C^3$; specifically,  photons with the helicity degree of freedom only (or electrons with spin only) are replaced by massive spin-one particles. Although such a `Free Will Experiment' has never been performed (though it might be, one day), \qm\ gives unambiguous predictions that may be used \emph{in lieu} of measurement outcomes. Compared to the set-up of the previous section, the following changes are to be made:
\begin{itemize}
\item The \emph{settings} are now given by $A=\ba$ and $B=\bb$, where
$\ba=[\vec{a}_1,\vec{a}_2,\vec{a}_3]$ and  $\bb=[\vec{b}_1,\vec{b}_2,\vec{b}_3]$
 are \emph{frames} in $\R^3$, that is, orthonormal bases $(\vec{a}_1,\vec{a}_2,\vec{a}_3)$ etc. in which each unit vector is defined up to a minus sign so that, e.g., $[-\vec{a}_1,\vec{a}_2,-\vec{a}_3]=[\vec{a}_1,\vec{a}_2,\vec{a}_3]$.
\item The \emph{outcomes} are now given by $F=\lm\in X_F$, $G=\mu\in X_G$, where
\begin{equation}
X_F=X_G=\{(1,1,0), (1,0,1), (0,1,1)\}. \label{XFXG}
\end{equation}
\item If we write $F=(F_1,F_2,F_3)$ and $G=(G_1,G_2,G_3)$, so that e.g.\ $F=(1,1,0)$ corresponds to $F_1=F_2=1, F_3=0$, the relevant outcome of the experiment in the \epr-state (defined in terms of the usual spin-1 basis $(|0\ra,|\pm 1\ra)$ of $\C^3$)
\beq|\psi_{\mbox{\sc epr}}\ra=(|-1\ra|-1\ra+|0\ra|0\ra+|1\ra|1\ra)/\sqrt{3},\eeq
at least as predicted by \qm,\footnote{What is being measured here by say Alice with setting $\ba$ is the triple
$(\la\vec{a}_1,\vec{J}\ra^2,\la\vec{a}_2,\vec{J}\ra^2,\la\vec{a}_3,\vec{J}\ra^2)$, where $\vec{J}$ is the angular momentum operator for spin one. Each operator $\la\vec{a}_i,\vec{J}\ra$ has spectrum $\{-1, 0,1\}$, so  each square $\la\vec{a}_i,\vec{J}\ra^2$ can be 0 or 1. Since $\vec{J}^2=2$, one has
$\la\vec{a}_1,\vec{J}\ra^2+\la\vec{a}_2,\vec{J}\ra^2+\la\vec{a}_3,\vec{J}\ra^2=2$, which gives \er{XFXG}.
}  is given by\footnote{
The complete (theoretical) statistics  are:  $P_{QM}(F_i=1,G_j=1|A=\ba,B=\bb) =\third(1+\la\vec{a}_i,\vec{b}_j\ra^2)$,\\
$P_{QM}(F_i=0,G_j=0|A=\ba,B=\bb)= \third\la\vec{a}_i,\vec{b}_j\ra^2$, 
$P_{QM}(F_i=1,G_j=0|A=\ba,B=\bb)=  \third(1-\la\vec{a}_i,\vec{b}_j\ra^2)$, and 
$P_{QM}(F_i=0,G_j=1|A=\ba,B=\bb)= \third(1-\la\vec{a}_i,\vec{b}_j\ra^2)$. See footnote \ref{fn7} for  notation like $P(F_i\neq G_j|\cdot)$.
}
\begin{equation}
P_{QM}(F_i\neq G_j|A=\ba,B=\bb)=\tthird \sin^2\theta_{\vec{a}_i,\vec{b}_j} \:\:\: (i,j=1,2,3). \label{uitkomstFWE} 
\end{equation}
Here $\theta_{\vec{a},\vec{b}}$ is the angle between $\vec{a}$ and $\vec{b}$, so that $\cos^2 \theta_{\vec{a},\vec{b}}=\la\vec{a},\vec{b}\ra^2$, cf.\ \er{uitkomstEPR}.
Note that the right-hand side only depends on $(\vec{a}_i,\vec{b}_j)$ rather than on all six vectors $(\ba,\bb)$.
\end{itemize}
Along the same lines as Theorem \ref{Bell1964}, and subject to analogous definitions,\footnote{See Definition \ref{FWTDEFS} below for Determinism, and Definition \ref{Belldefs} for the others, \emph{mutatis mutandis}.} one  proves:
\begin{Theorem}\label{BellFWT}
Determinism, Probability Theory,  Parameter Independence, Freedom,  and Nature (i.e.\ the outcome \er{uitkomstFWE} of the Free Will Experiment) are contradictory.
\end{Theorem}

 For future reference, we also record the following consequence of \er{uitkomstFWE}:\footnote{Here $P_{QM}(F_i=G_j|A_i=B_j)$ denotes $P_{QM}(F_i=0,G_j=0|A_i=B_j)+P_{QM}(F_i=1,G_j=1|A_i=B_j)$, where the setting $A_i=B_j$ stands for $(A=\ba,B=\bb)$ subject to $\vec{a}_i=\pm\vec{b}_j$. It follows from \er{uitkomstFWE} or the previous footnote that $P_{QM}(F_i=G_j|A_i=B_j)=\third(1+2\cos^2 \theta_{\vec{a}_i,\vec{b}_j})$, which for $\vec{a}_i=\pm\vec{b}_j$ equals unity.}
\begin{equation}
P_{QM}(F_i=G_j|A_i=B_j)=1.\label{CKpc}
\end{equation}
In other words, if the settings $(\ba,\bb)$ have $\vec{a}_i=\pm\vec{b}_j$, then  with probability one the measurements $F_i$ and $G_j$ have the same outcomes (i.e.\ either $F_i=G_j=0$ or $F_i=G_j=1$).
\section{The Strong Free Will Theorem revisited}\label{FWTsec}
The Strong Free Will Theorem \cite{CK2} historically arose as a refinement of the Kochen--Specker Theorem \cite{Bell2,KS}, in which the assumption of \emph{Non-contextuality} in a single-wing experiment on a (massive) spin-one particle was replaced by the assumption of Parameter Independence in the double-wing experiment described in the previous section. In turn, the Kochen--Specker Theorem (like Gleason's Theorem, from which it follows) freed von Neumann's no-go result for hidden variable theories \cite{vN32} from its controversial linearity assumption (see \cite{Bub2} for a balanced discussion). Thus the Strong Free Will Theorem of 2009  may be seen as a finishing touch of the development started by von Neumann in 1932.
 Ironically, we are now going to place the  Strong Free Will Theorem in the Bell tradition, which emphatically arose in opposition (if not hostility) to the work of von Neumann! 
 
 Roughly speaking, the Strong Free Will Theorem removes the assumption of Probability Theory from Bell's (1964) Theorem (in our spin-one version, i.e., Theorem \ref{BellFWT}), but in order to achieve this, some of the assumptions now acquire a somewhat different meaning.

\begin{Definition}\label{FWTDEFS}
In the context of the Free Will Experiment of the previous section:
\begin{itemize}
\item 
\textbf{Determinism} means that there is a state space $X$  with associated functions $$A: X\raw X_A, B:X\raw X_B, F:X\raw X_F, G:X\raw X_G,$$ where $X_A=X_B$ is  the set of all frames in $\R^3$, and $X_F=X_G$ is given by \er{XFXG},
which completely describe the experiment in the sense that each state $x\in X$ determines both its settings $\ba=A(x),\bb=B(x)$ and its outcome $\lm=F(x),\mu=G(x)$. 

Furthermore, in terms of a postulated additional random variable $Z:X\raw X_Z$:
\item \textbf{Parameter Independence}  means that 
$F=F(A,Z)$ and $G=G(B,Z)$,
 i.e., for all $x\in X$ one has $F(x)=\hat{F}(A(x),Z(x))$ and $G(x)=\hat{G}(B(x),Z(x))$ for certain functions
  $\hat{F}:X_A \x X_Z\raw X_F$, $\hat{G}:X_B\x X_Z\raw X_G$.
\item \textbf{Freedom} means that
 $(A,B,Z)$ are independent in the sense that for each $(\ba,\bb,z)\in X_A\x X_B\x X_Z$ there is an $x\in X$  for which $A(x)=\ba$, $B(x)=\bb$, and $Z(x)=z$.
\end{itemize}
\end{Definition}
Thus the main change lies in the Freedom assumption, which simply says that the function $A\x B\x Z:X\raw X_A\x X_B\x X_Z, x\mapsto (A(x),B(x),Z(x))$, is surjective. The goal of this assumption is to remove any potential  dependencies between (or constraints on)  the variables  $(\ba,\bb,z)$, and hence between the physical system Alice and Bob perform their measurements \emph{on}, and the devices they perform their measurements \emph{with}.

Also, rather than the probabilistic outcome  \er{uitkomstFWE} of the Free Will Experiment, we use its corollary \er{CKpc}, construed non-probabilistically (i.e., probability one is replaced by deterministic certainty):  writing $\hat{F}=(\hat{F}_1,\hat{F}_2,\hat{F}_3)$ and  $\hat{G}=(\hat{G}_1,\hat{G}_2,\hat{G}_3)$, analogous to $F$ and $G$, so that $\hat{F}_i:X_A \x X_Z\raw \{0,1\}$ and $\hat{G}_j:X_B \x X_Z\raw \{0,1\}$, \emph{Nature} reveals that:\footnote{To keep matters simple, we will not be bothered with the notational difference between frames $[\vec{a}_1,\vec{a}_2,\vec{a}_3]$ and orthonormal bases $(\vec{a}_1,\vec{a}_2,\vec{a}_3)$, and similarly for $\bb$, until the end of the proof.}
 \begin{equation}
\vec{a}_i=\vec{b}_j\: \Raw\: \hat{F}_i(\vec{a}_1,\vec{a}_2,\vec{a}_3,z)=\hat{G}_j(\vec{b}_1,\vec{b}_2,\vec{b}_3,z). \label{ienj}
\end{equation}
Our reformulation of the Strong Free Will Theorem \cite{BS,Clifton,CK2,HR,Stairs}, then, is as follows.
\begin{Theorem}\label{CKFWT}
Determinism, Parameter Independence, Freedom,  and Nature (here represented by the outcome \er{ienj} of the Free Will Experiment) are  contradictory.
\end{Theorem}
\newpage
\emph{Proof.} 
The Freedom assumption allows us to treat $(\ba,\bb,z)$ as free variables, a fact that will tacitly be used all the time.
First, take $i=j$ in \er{ienj}. This shows that  $\hat{F}_i(\vec{a}_1,\vec{a}_2,\vec{a}_3,z)$ only depends on $(\vec{a}_i,z)$, whilst
$\hat{G}_j(\vec{b}_1,\vec{b}_2,\vec{b}_3,z)$ only depends on $(\vec{b}_j,z)$. Hence we write $\hat{F}_i(\vec{a}_1,\vec{a}_2,\vec{a}_3,z)=\til{F}_i(\vec{a}_i,z)$, etc. 
Next, taking 
 $i\neq j$ in \er{ienj} shows that $\til{F}_1(\vec{a},z)=\til{F}_2(\vec{a},z)=\til{F}_3(\vec{a},z)$.
Consequently, the function  $\hat{F}:X_A \x X_Z\raw X_F$ is given by
\begin{equation}
\hat{F}(\vec{a}_1,\vec{a}_2,\vec{a}_3,z)=(\til{F}(\vec{a}_1,z),\til{F}(\vec{a}_2,z),\til{F}(\vec{a}_3,z)),
\end{equation}
Combined with its value set \er{XFXG}, this
shows that for each fixed $z$, $\hat{F}$ is a \emph{frame function}: to each frame $\ba$ it assigns one of the triples in \er{XFXG}, in such a way that if two different frames $\ba$ and $\ba'$ overlap in that $\vec{a}'_i=\vec{a}_j$ for some $i,j$, then $\hat{F}_i(\vec{a}'_i,z)=\hat{F}_j(\vec{a}_j,z)$.
However, such a function does not exist by the Kochen--Specker Theorem \cite{KS,Peres}.\enp

\noindent Through the proof of the Kochen--Specker Theorem, this proof  shows that a suitable finite set of frames will do for $X_A=X_B$, a simplification that is not available in Theorem \ref{BellFWT}!
\section{Bell's  Theorem revisited} \label{BLTsec}
To close, we show that what is usually called Bell's Theorem \cite{Bell4,Bub,JB,Gill,Jaeger,Maudlin,Seevinck,WW}, in which \emph{Determinism} is not assumed, may also be reformulated using our  treatment of  apparatus settings as random variables.
We restrict ourselves to generalizing Theorem \ref{Bell1964};  Theorem \ref{BellFWT} may be adapted to stochastic hidden variables in an  analogous way.
\begin{Definition}\label{Bellnewdefs}
In the context of the  \epr-Bohm experiment (with photons):
\begin{itemize}
\item
\textbf{Probability Theory} means that there is a probability space $(X,\Sg,P)$, carrying random variables 
\er{ABFG}, so that the empirical  probabilities are reproduced as conditional joint probabilities through \er{PEP}.
\item \textbf{Bell-Locality}  means that there is a fifth random variable $Z:X\raw X_Z$ for which
\begin{eqnarray} && P(F=\lm,G=\mu|A=\al,B=\beta,Z=z)=  \\
&& P(F=\lm|A=\al,Z=z)\cdot P(G=\mu|B=\beta,Z=z).\label{BellL}
\end{eqnarray}
\item \textbf{Freedom} means that, for this fifth variable, $P(Z=z|A=\al,B=\beta)=P(Z=z)$.
\end{itemize}
\end{Definition}
\begin{Theorem}\label{Bell1976}
 Probability Theory,  Bell-Locality, Freedom, and Nature are contradictory, where Nature  is represented through the outcome \er{uitkomstEPR} of the \epr-Bohm experiment.
\end{Theorem}
\emph{Proof.} Introduce a new probability space $\tilde{X}_Z=[0,1]\x [0,1]\x X_Z$, with elements  $(s,t,z)$, and  probability measure $d\til{P}_Z(s,t,z)=ds\cdot dt\cdot dP_Z(z)$. On $\tilde{X}_Z$, define random variables
\begin{eqnarray}
\til{F}_{\al}(s,t,z)&=& \chi_{[0,P(F=1|A=\al,Z=z)]}(s);\\
\til{G}_{\beta}(s,t,z)&=& \chi_{[0,P(G=1|B=\beta,Z=z)]}(t),
\end{eqnarray}
a move inspired by \cite{WW}. Using all assumptions of the theorem, one then finds
\begin{equation}
P(F=\lm,G=\mu|A=\al,B=\beta)=\til{P}_Z(\til{F}_{\al}=\lm,\til{G}_{\beta}=\mu), \label{crux3bis}
\end{equation}
cf.\ \er{crux2bis}, so that the proof may be completed exactly as in the case of Theorem \ref{Bell1964}. \enp

    \section*{Acknowledgement}
The authors are indebted to Jeff Bub, Jeremy Butterfield, Dennis Dieks, Richard Gill, Hans Maassen, and Matt Leifer for various comments on this work (including  predecessors).
\end{document}